\documentclass[twocolumn]{aastex62}
\usepackage{lineno}
\usepackage{footnote}
\usepackage{amsmath}
\usepackage{amsfonts}
\usepackage{graphicx,latexsym,amssymb,epsfig}
\usepackage{xcolor}

\begin{document}
\title{Hadron-quark Pasta Phase in Massive Neutron Stars}

\author[0000-0002-4467-8521]{Min Ju}
\affiliation{School of Physics, Nankai University, Tianjin 300071, China}

\author[0000-0002-1709-0159]{Jinniu Hu}
\affiliation{School of Physics, Nankai University, Tianjin 300071, China; hujinniu@nankai.edu.cn}

\author[0000-0003-2717-9939]{Hong Shen}
\affiliation{School of Physics, Nankai University, Tianjin 300071, China; shennankai@gmail.com}

\begin{abstract}
The structured hadron-quark mixed phase, known as the pasta phase,
is expected to appear in the core of massive neutron stars.
Motivated by the recent advances in astrophysical observations,
we explore the possibility of the appearance of quarks inside neutron stars
and check its compatibility with current constraints.
We investigate the properties of the hadron-quark pasta phases and
their influences on the equation of state (EOS) for neutron stars.
In this work, we extend the energy minimization (EM) method to describe the
hadron-quark pasta phase, where the surface and Coulomb contributions are
included in the minimization procedure. By allowing different electron densities
in the hadronic and quark matter phases, the total electron chemical potential
with the electric potential remains constant, and local $\beta$ equilibrium is
achieved inside the Wigner--Seitz cell.
The mixed phase described in the EM method shows the features lying between
the Gibbs and Maxwell constructions, which is helpful for understanding the
transition from the Gibbs construction (GC) to the Maxwell construction (MC) with
increasing surface tension.
We employ the relativistic mean-field model to describe the hadronic matter,
while the quark matter is described by the MIT bag model with vector interactions.
It is found that the vector interactions among quarks can significantly
stiffen the EOS at high densities and help enhance the maximum mass of neutron stars.
Other parameters like the bag constant can also affect
the deconfinement phase transition in neutron stars.
Our results show that hadron-quark pasta phases may appear in the core
of massive neutron stars that can be compatible with current observational
constraints.
\end{abstract}

\keywords{Neutron stars --- Nuclear astrophysics --- neutron star cores ---  gravitational waves}

\section{I\lowercase{ntroduction}}
\label{sec:1}
The appearance of deconfined quark matter, which is expected to appear in the core of
massive neutron stars, has received increasing attention recently because
of its relevance to astrophysical observations~\citep{Latt16,Baym18,Anna20}.
In the last decade, several breakthrough discoveries in astronomy provided
valuable information and constraints on the properties of neutron stars.
The precise mass measurements of
PSR J1614-2230 ($1.908 \pm 0.016   M_\odot$;~\citeauthor{Demo10} \citeyear{Demo10};
                                              \citeauthor{Fons16} \citeyear{Fons16};
                                              \citeauthor{Arzo18} \citeyear{Arzo18}),
PSR J0348+0432  ($2.01  \pm 0.04   M_\odot$;~\citeauthor{Anto13} \citeyear{Anto13}), and
PSR J0740+6620  ($2.08  \pm 0.07   M_\odot$;~\citeauthor{Crom20} \citeyear{Crom20};
                                              \citeauthor{Fons21} \citeyear{Fons21})
constrain the maximum neutron-star mass $M_{\rm{max}}$ to be larger than
about $2 M_\odot$, which poses a challenge to our understanding of the
equation of state (EOS) of superdense matter.
The recent observations by the Neutron Star Interior Composition Explorer (NICER)
provided a simultaneous measurement of the mass and radius for
PSR J0030+0451, which was reported to have a mass of
$1.44_{-0.14}^{+0.15} M_\odot$ with a radius of $13.02_{-1.06}^{+1.24}$ km~\citep{Mill19}
and a mass of $1.34_{-0.16}^{+0.15} M_\odot$ with a radius
of $12.71_{-1.19}^{+1.14}$ km~\citep{Rile19} by two independent groups.
The new measurements by NICER for the most massive known neutron star,
PSR J0740+6620 ($2.08 \pm 0.07  M_\odot$),
showed that it has a radius of $13.7_{-1.5}^{+2.6}$~km by~\citet{Mill21}
and $12.39_{-0.98}^{+1.30}$ km by~\citet{Rile21}.
In particular, the discovery of gravitational waves from a binary neutron-star merger
event GW170817 has opened a new era of multimessenger astronomy~\citep{Abbo17}.
Based on the observations of GW170817, the tidal deformability of a canonical $1.4 M_\odot$
neutron star was estimated to be $70<\Lambda_{1.4}<580$ and the corresponding
radius was inferred to be $10.5<R_{1.4}<13.3$~km~\citep{Abbo18}.
More recently, the gravitational-wave events, GW190425~\citep{Abbo190425} and
GW190814~\citep{Abbo190814}, were reported by LIGO and Virgo Collaborations.
The total mass of the GW190425 system is as large as $3.4^{+0.3}_{-0.1} M_\odot$,
which is even more massive than any neutron-star binary observed so far;
hence its gravitational-wave analyses may offer valuable information for the EOS
at high densities and possible phase transitions inside neutron stars.
The GW190814 event was detected from a compact binary coalescence involving a
22.2--24.3$M_\odot$ black hole and a 2.50--2.67$M_\odot$ compact object
that could be either the heaviest neutron star
or the lightest black hole ever observed~\citep{Abbo190814}.
The discovery of GW190814 has triggered many theoretical efforts exploring
the nature of the secondary object and its implications for high-density
EOS~\citep{Fatt20,Huang20,Tews21}.
The gravitational-wave analyses~\citep{Essi20,Tews21} suggested that
the secondary in GW190814 is more likely to be a black hole, but
its possibility as a neutron star cannot be ruled out.
Several studies have suggested that the secondary in GW190814 may be a rapidly
rotating neutron star~\citep{Li20,Most20,Tsok20,Zhang20}.
Moreover, it could be also considered as a heavy neutron star containing deconfined
quark matter, where the inclusion of quarks has significant impact on
the observations~\citep{Tan20,Demi21,Dexh21}.
In light of these recent developments in astronomy, it would be
interesting and informative to explore possible structures of
the hadron-quark mixed phase in massive neutron stars.

For the description of hadron-quark phase transition in neutron stars,
both Gibbs and Maxwell constructions are often used depending on the
surface tension at the interface~\citep{Bhat10}.
In the limit of zero surface tension, the Gibbs equilibrium conditions
are satisfied between the two coexisting
phases, and only global charge neutrality is imposed in the mixed
phase~\citep{Glen92,Sche00,Yang08,Xu10,Wu17}.
On the other hand, the Maxwell construction (MC) is valid for sufficiently large
surface tension, where local charge neutrality is enforced and the transition
takes place at constant pressure~\citep{Bhat10,Han19,Wu19}.
It is noteworthy that only bulk contributions
are involved in the Gibbs and Maxwell constructions, whereas the finite-size
effects like surface and Coulomb energies are neglected.
In a more realistic case, due to the competition between surface and Coulomb
energies, some geometric structures may be formed in the hadron-quark mixed phase,
known as hadron-quark pasta phases~\citep{Heis93,Endo06,Maru07,Yasu14,Spin16,Webe19,Wu19}.
This structured mixed phase is analogous
to nuclear pasta phase in the inner crust of neutron stars.
Several methods have been developed to study the properties of hadron-quark pasta
phases. In the coexisting phases (CP) method~\citep{Wu19}, the hadronic and quark
phases are assumed to satisfy the Gibbs conditions for phase equilibrium,
while the surface and Coulomb energies are taken into account perturbatively.
A more realistic description of the pasta phase has been developed in a series of
works~\citep{Endo06,Maru07,Yasu14}, where the Thomas--Fermi approximation was used to
describe the density profiles of hadrons and quarks in the Wigner--Seitz cell.
For simplicity, the particle densities in the two coexisting phases are generally
assumed to be spatially constant, and the charge screening effect is neglected.
In our previous works~\citep{Wu19,Ju21}, we proposed an energy minimization (EM) method
for improving the treatment of surface and Coulomb energies, which play a key role in
determining the structure of the pasta phases. By incorporating the surface and Coulomb
contributions in the EM procedure, one can derive the equilibrium
conditions for coexisting phases that are different from the
Gibbs conditions used in the CP method.
In the present work, we further extend the EM method for exploring the hadron-quark
pasta phases in massive neutron stars.

In the EM method, the hadron-quark pasta phase is described within the Wigner--Seitz
approximation, where the whole space is divided into equivalent cells with a
geometric symmetry. The hadronic and quark phases in a charge-neutral cell are
assumed to be separated by a sharp interface. In the more realistic description
of~\citet{Tats03}, there is a thin boundary layer between two separate bulk phases,
which can lead to a difference in the electric potential between the two phases.
As a result, the electron densities in the hadronic and quark phases are allowed
to be different from each other, whereas the electron chemical potentials are equal.
According to this argument, the MC could be stable although the
electron densities in two separate bulk phases are different.
The difference in the electron densities between two separate phases can be
understood as a result of the charge screening effect.
In the present study, we would like to evaluate how important the charge screening
effect is for the hadron-quark pasta phases. Meanwhile, the results in the EM method
will be compared to those obtained with a uniform electron gas in the CP method.

To describe the hadronic phase, we employ the relativistic mean-field (RMF) model
and choose the recently proposed BigApple parameterization~\citep{Fatt20}.
In the past decades, the RMF approach based on various energy density functionals
has been successfully applied in the description of finite nuclei and infinite nuclear
matter. Several popular RMF models, such as NL3~\citep{Lala97}, TM1~\citep{Suga94},
and IUFSU~\citep{Fatt10}, have been widely used in astrophysical applications,
since they can not only reproduce experimental data of finite nuclei
but also predict large enough neutron-star masses.
The BigApple model was proposed by~\citet{Fatt20} after the discovery of GW190814,
in which they considered various constraints from astrophysical observations as well as
the ground-state properties of finite nuclei.
The BigApple model predicts a maximum neutron-star mass of $M_{\rm{max}}=2.6 M_\odot$,
while the resulting radius and tidal deformability of neutron stars are consistent
with GW170817 and NICER observations.
It is well known that nuclear symmetry energy and its slope play
an important role in understanding the properties of neutron stars~\citep{Oert17,Ji19}.
There exists a positive correlation between the symmetry energy slope $L$ and the
neutron-star radius~\citep{Alam16}.
The NL3 and TM1 models have a rather large slope parameter, which results in too large
a radius and tidal deformability for a canonical $1.4 M_\odot$ neutron star
as compared to the estimations from astrophysical observations~\citep{Ji19}.
In the present work, we prefer to employ the BigApple model with a small slope
$L$, which is more consistent with the analysis of the GW170817 event.
Generally, the appearance of hyperons at high densities would considerably soften
the EOS and reduce the maximum neutron-star mass~\citep{Oert17}.
However, the influence of hyperons may be suppressed by incorporating quark degrees
of freedom or introducing additional repulsion for hyperons~\citep{Lona15,Gome19}.
Currently, there are large uncertainties in the hyperon--nucleon and hyperon--hyperon
interactions due to limited experimental data.
For simplicity, we do not include hyperons in the present calculations
and focus on the influence of deconfined quarks in the core of neutron stars.

We utilize a modified MIT bag model with vector interactions, often referred to as
the vMIT model~\citep{Gome19,Han19}, to describe the quark phase.
It has been shown in the literature that including vector interactions among quarks
can significantly stiffen the EOS at high densities and help to produce massive
neutron stars in agreement with the astronomical observations~\citep{Klah15}.
The vector interaction in the vMIT model is introduced via the exchange of a vector meson
that is analogous to the $\omega$ meson in the Walecka model~\citep{Lope21}.
The vector coupling constant is usually treated as a free parameter
and a universal coupling is assumed for all quark flavors.
In~\citet{Han19} and \citet{Gome19}, the influences of the vector interaction on the properties of compact stars
were studied using the Gibbs and Maxwell constructions to describe the hadron-quark mixed phase
without considering possible geometric structures.
In the present work, we intend to investigate how the hadron-quark pasta phases are
affected by the vector interactions among quarks within the EM method.

We have two aims in this paper. The first is to explore the possibility of
the appearance of hadron-quark pasta phases in the core of massive neutron stars.
By comparing with current constraints, we investigate the compatibility between
the hadron-quark phase transition and astrophysical observations,
as well as the influence of the quark vector interactions.
The second aim is to extend the EM method for describing the hadron-quark
pasta phases with the charge screening effect.
By allowing different electron densities in separate bulk phases,
the local $\beta$ equilibrium is achieved inside the Wigner--Seitz cell.
Although in more precise Thomas--Fermi calculations the electron density
should depend on the position, a simplified treatment of the charge screening
effect with different constant electron densities in the two bulk phases
is helpful for understanding the transition from the GC
to the MC.
Comparing with the GC, the EM method used in this work
involves the effects of finite size and charge screening,
which are essential for determining the structure of the hadron-quark mixed phase.

This paper is arranged as follows.
In Section~\ref{sec:2}, we briefly describe the framework for describing
the hadronic and quark phases.
The EM method for the hadron-quark pasta phases is presented in Section~\ref{sec:3}.
In Section~\ref{sec:4}, we discuss numerical results of the hadron-quark pasta phases
and neutron-star properties.
Section~\ref{sec:5} is devoted to a summary.

\section{M\lowercase{odels for} h\lowercase{adronic and} q\lowercase{uark} p\lowercase{hases}}
\label{sec:2}

In this section, we briefly describe the RMF model for the hadronic matter and
the vMIT model for the quark matter. In addition, we explain the model parameters
used in our calculations.

\subsection{Hadronic Phase}
\label{sec:2.1}

We employ the RMF model with the BigApple parameterization to describe
the hadronic matter, where nucleons interact through the exchange of various mesons
including the isoscalar--scalar $\sigma$ meson, the isoscalar--vector $\omega$ meson,
and the isovector--vector $\rho$ meson.
The Lagrangian density for hadronic matter consisting of nucleons ($p$ and $n$) and
leptons ($e$ and $\mu$) is written as
{\small
\begin{eqnarray}
\label{eq:LRMF}
\mathcal{L} & = & \sum_{i=p,n}\bar{\psi}_i
\left\{i\gamma_{\mu}\partial^{\mu}-M-g_{\sigma}\sigma
-\gamma_{\mu} \left[g_{\omega}\omega^{\mu} +\frac{g_{\rho}}{2}\tau_a\rho^{a\mu}
\right]\right\}\psi_i  \nonumber \\
&& +\frac{1}{2}\partial_{\mu}\sigma\partial^{\mu}\sigma -\frac{1}{2}%
m^2_{\sigma}\sigma^2-\frac{1}{3}g_{2}\sigma^{3} -\frac{1}{4}g_{3}\sigma^{4}
\nonumber \\
&& -\frac{1}{4}W_{\mu\nu}W^{\mu\nu} +\frac{1}{2}m^2_{\omega}\omega_{\mu}%
\omega^{\mu} +\frac{1}{4}c_{3}\left(\omega_{\mu}\omega^{\mu}\right)^2  \nonumber \\
&& -\frac{1}{4}R^a_{\mu\nu}R^{a\mu\nu} +\frac{1}{2}m^2_{\rho}\rho^a_{\mu}%
\rho^{a\mu}
+\Lambda_{\rm{v}} \left(g_{\omega}^2
\omega_{\mu}\omega^{\mu}\right)
\left(g_{\rho}^2\rho^a_{\mu}\rho^{a\mu}\right) \nonumber\\
&& +\sum_{l=e,\mu}\bar{\psi}_{l}
  \left( i\gamma_{\mu }\partial^{\mu }-m_{l}\right)\psi_l,
\end{eqnarray} }
where $W^{\mu\nu}$ and $R^{a\mu\nu}$ represent the antisymmetric field
tensors for $\omega^{\mu}$ and $\rho^{a\mu}$, respectively.
Under the mean-field approximation, the meson fields are treated as classical
fields, which are denoted by
$\sigma =\left\langle \sigma \right\rangle$,
$\omega =\left\langle\omega^{0}\right\rangle$,
and $\rho =\left\langle \rho^{30} \right\rangle$.
These mean fields can be obtained by solving a set of coupled equations in the RMF model.

For the hadronic matter in $\beta$ equilibrium, the chemical potentials
satisfy the relations $\mu_{p}=\mu_{n}-\mu_{e}$ and $\mu_{\mu}=\mu_{e}$.
At zero temperature, the chemical potentials of nucleons and leptons are expressed as
\begin{eqnarray}
\mu_i &=& \sqrt{{k_{F}^{i}}^{2}+{M^{\ast}}^2}+g_{\omega}\omega
+\frac{g_{\rho}}{2}\tau_{3} \rho, \hspace{0.5cm}  i=p,n,
\label{eq:mup}\\
\mu_{l} &=& \sqrt{{k_{F}^{l}}^{2}+m_{l}^{2}}, \hspace{3.0cm}  l=e,\mu,
\label{eq:mul}
\end{eqnarray}
with $\tau_{3}=+1$ and $-1$ for protons and neutrons, respectively.
The effective nucleon mass is defined as $M^{\ast}=M+g_{\sigma}{\sigma}$.
The total energy density and pressure in hadronic matter are written as
\begin{eqnarray}
\varepsilon_{\rm{HP}} &=&\sum_{i=p,n}  \varepsilon^i_{\rm{FG}}
        +\sum_{l=e,\mu} \varepsilon^l_{\rm{FG}} \nonumber \\
&&   + \frac{1}{2}m^2_{\sigma}{\sigma}^2+\frac{1}{3}{g_2}{\sigma}^3
     + \frac{1}{4}{g_3}{\sigma}^4  + \frac{1}{2}m^2_{\omega}{\omega}^2 \nonumber  \\
&&   + \frac{3}{4}{c_3}{\omega}^4
     + \frac{1}{2}m^2_{\rho}{\rho}^2
     + 3{\Lambda}_{\rm{v}}\left(g^2_{\omega}{\omega}^2\right)
     \left(g^2_{\rho}{\rho}^2\right) ,
     \label{eq:ehp} \\
P_{\rm{HP}} &=&\sum_{i=p,n}  P^i_{\rm{FG}}
        +\sum_{l=e,\mu} P^l_{\rm{FG}} \nonumber \\
&&   - \frac{1}{2}m^2_{\sigma}{\sigma}^2-\frac{1}{3}{g_2}{\sigma}^3
     - \frac{1}{4}{g_3}{\sigma}^4  + \frac{1}{2}m^2_{\omega}{\omega}^2 \nonumber \\
&&   + \frac{1}{4}{c_3}{\omega}^4
     + \frac{1}{2}m^2_{\rho}{\rho}^2
     + \Lambda_{\rm{v}}\left(g^2_{\omega}{\omega}^2\right)
     \left(g^2_{\rho}{\rho}^2\right) ,
     \label{eq:php}
\end{eqnarray}
where $\varepsilon^i_{\rm{FG}}$ and $P^i_{\rm{FG}}$ denote the Fermi gas contributions
of species $i$ with a mass $m_i$ and degeneracy $N_i$,
\begin{eqnarray}
\varepsilon^i_{\rm{FG}} &=& N_i
     \int_{0}^{k^{i}_{F}} \frac{d^{3}k}{(2\pi)^3} \sqrt{k^2+m_i^2} ,
     \label{eq:efg}   \\
P^i_{\rm{FG}} &=& \frac{N_i}{3}
     \int_{0}^{k^{i}_{F}} \frac{d^{3}k}{(2\pi)^3} \frac{k^2}{\sqrt{k^2+m_i^2}} .
     \label{eq:pfg}
\end{eqnarray}
For nucleons, the effective mass $m_i=M^{\ast}$ is used and $N_i=2$ represents the spin degeneracy.

In the present calculations, we employ the BigApple parameterization given in~\citet{Fatt20},
which could provide an accurate description of ground-state properties for finite nuclei
across the nuclear chart. Its predictions for infinite nuclear matter at the saturation density
$n_0=0.155\,\rm{fm}^{-3}$ are
$E_0=-16.34\,\rm{MeV}$ for energy per nucleon,
$K=227\,\rm{MeV}$ for incompressibility,
$E_{\rm{sym}}=31.3\,\rm{MeV}$ for symmetry energy,
and $L=39.8\,\rm{MeV}$ for the slope of symmetry energy.
The small slope parameter $L$ in the BigApple model leads to acceptable radius
and tidal deformability for a canonical $1.4 M_\odot$ neutron star,
as compared to the estimations from astrophysical observations.
Moreover, the BigApple model predicts a maximum neutron-star mass of $2.6 M_\odot$,
which is sufficiently large for exploring possible phase transitions in neutron stars.

\subsection{Quark Phase}
\label{sec:2.2}

We use a modified MIT bag model with vector interactions (vMIT) to describe the quark phase.
The inclusion of repulsive vector interactions among quarks plays
a crucial role in obtaining a stiff high-density EOS that is required by
the observations of $\sim 2 M_\odot$ neutron stars.
In the vMIT model, the vector interaction is introduced via the exchange of
a vector meson with the mass $m_V$. We consider the quark matter consisting
of three flavor quarks ($u$, $d$, and $s$) and leptons ($e$ and $\mu$).
The Lagrangian density of the vMIT model in the mean-field approximation is written as
\begin{eqnarray}
\label{eq:LvMIT}
\mathcal{L} &=& \sum_{i=u,d,s}\left[\bar{\psi}_{i}\left( i\gamma_{\mu}\partial^{\mu}
             -m_{i}-g_{V}\gamma_{\mu}V^{\mu}\right)\psi_{i} -B
             \right. \nonumber  \\
            & & \left. +\frac{1}{2}m_{V}^{2}V_{\mu}V^{\mu}  \right]\Theta
             +\sum_{l=e,\mu}\bar{\psi_{l}}\left( i\gamma_{\mu}\partial^{\mu}
             -m_{l}\right)\psi_{l},
\end{eqnarray}
where $B$ denotes the bag constant and $\Theta$ is the Heaviside step function
representing the confinement of quarks inside the bag.
The nonzero mean field $V_0$ is calculated from the equation of motion for the vector meson,
\begin{eqnarray}
\label{eq:eqv0}
m_{V}^{2}V_{0} &=& g_{V} \sum_{i=u,d,s} n_i ,
\end{eqnarray}
with $n_i$ being the number density of the quark flavor $i$.
The quark chemical potential is then given by
\begin{eqnarray}
\label{eq:muq}
\mu_i=\sqrt{{k^{i}_{F}}^{2}+m_{i}^{2}}+g_{V}V_{0},
\end{eqnarray}
which is clearly enhanced by the vector potential.

The total energy density and pressure in quark matter are written as
\begin{eqnarray}
\varepsilon_{\rm{QP}} &=& \sum_{i=u,d,s} \varepsilon^i_{\rm{FG}}
        +\sum_{l=e,\mu} \varepsilon^l_{\rm{FG}} \nonumber \\
&&      +\frac{1}{2} \left( \frac{g_V}{m_V}\right)^2 \left(n_u+n_d+n_s\right)^{2} +B ,
\label{eq:eqp}\\
P_{\rm{QP}} &=& \sum_{i=u,d,s} P^i_{\rm{FG}}
        +\sum_{l=e,\mu} P^l_{\rm{FG}}  \nonumber \\
&&      +\frac{1}{2} \left( \frac{g_V}{m_V}\right)^2 \left(n_u+n_d+n_s\right)^{2} -B ,
\label{eq:pqp}
\end{eqnarray}
where $\varepsilon^i_{\rm{FG}}$ and $P^i_{\rm{FG}}$ denote the Fermi gas contributions
of species $i$, as given by Equations~(\ref{eq:efg}) and~(\ref{eq:pfg}).
For the quark flavor $i$, the degeneracy $N_i=6$ arises from the spin and color degrees
of freedom, while $m_i$ represents the current quark mass.
The vector interactions among quarks are crucial for high-density EOS.
In practice, we vary the vector coupling $G_V=\left( g_V/m_V \right)^2$
in the range of 0--0.3 fm$^2$ in order to examine the impact of vector interactions.
We adopt the current quark masses $m_{u}=m_{d}=5.5$ MeV and $m_{s}=95$ MeV in our
calculations. As for the bag constant, we mainly use the value of $B^{1/4}=180$~MeV.
It is well known that the bag constant could significantly affect the EOS of quark matter
and consequently influence the hadron-quark phase transition.
We will compare the results with different choices of $B$ in the vMIT model for quark matter,
to evaluate the influence of the bag constant on the hadron-quark
pasta phases in neutron stars.

\section{H\lowercase{adron-quark} p\lowercase{asta} p\lowercase{hases}}
\label{sec:3}

To describe the hadron-quark pasta phases, we employ the (EM) method
within the Wigner--Seitz approximation, where the whole space is divided into equivalent
cells with a geometric symmetry.
The coexisting hadronic and quark phases in a charge-neutral cell are assumed to be
separated by a sharp interface, while the particle densities in each phase are taken
to be uniform for simplicity. This is analogous to the compressible liquid-drop
model used in the study of nuclear liquid-gas phase transition at subnuclear
densities~\citep{Latt91,Bao14}.
It is well known that the surface tension at the interface plays a crucial role
in determining the structure of the mixed phase.
The Gibbs and Maxwell constructions, respectively, correspond to
the two extreme cases of zero and large surface tension.
The hadron-quark mixed phase described in the EM method can be understood as an
intermediate state lying between the Gibbs and Maxwell constructions.
For the EM method used in our previous works~\citep{Wu19,Ju21}, the electrons
are assumed to be uniformly distributed throughout the whole cell.
This is consistent with the GC, but contradicts the MC
where the requirement of local charge neutrality leads to different
electron densities in the two phases.
In order to understand the transition from the GC to the MC
with increasing surface tension, we extend the EM method by allowing
different electron densities in the hadronic and quark phases in the present study.
This may be caused by the difference in the electric potential between the two phases,
considered as a result of the charge screening effect.

In the EM method, the total energy density of the mixed phase is expressed as
\begin{eqnarray}
\label{eq:fws}
\varepsilon_{\rm{MP}} &=& \chi \varepsilon_{\rm{QP}}
  + \left( 1 - \chi \right)\varepsilon_{\rm{HP}}
  + \varepsilon_{\rm{surf}} + \varepsilon_{\rm{Coul}} ,
\end{eqnarray}%
where $\chi=V_{\rm{QP}}/(V_{\rm{QP}}+V_{\rm{HP}})$ denotes the volume fraction
of the quark phase. The energy densities, $\varepsilon_{\rm{HP}}$ and $\varepsilon_{\rm{QP}}$,
are given by Equations~(\ref{eq:ehp}) and (\ref{eq:eqp}), respectively.
The first two terms of Equation~(\ref{eq:fws}) represent the bulk contributions,
while the last two terms come from the finite-size effects.
The surface and Coulomb energy densities are calculated from
\begin{eqnarray}
{\varepsilon}_{\rm{surf}}
&=& \frac{D \sigma \chi_{\rm{in}}}{r_D},
\label{eq:esurf} \\
{\varepsilon}_{\rm{Coul}}
&=& \frac{e^2}{2}\left(\delta n_c\right)^{2}r_D^{2} \chi_{\rm{in}}\Phi\left(\chi_{\rm{in}}\right),
\label{eq:ecoul}
\end{eqnarray}%
with%
\begin{eqnarray}
\label{eq:Du}
\Phi\left(\chi_{\rm{in}}\right)=\left\{
\begin{array}{ll}
\frac{1}{D+2}\left(\frac{2-D\chi_{\rm{in}}^{1-2/D}}{D-2}+\chi_{\rm{in}}\right),  & D=1,3, \\
\frac{\chi_{\rm{in}}-1-\ln{\chi_{\rm{in}}}}{D+2},  & D=2, \\
\end{array} \right.
\end{eqnarray}%
where $D=1,2,3$ denotes the geometric dimension of the cell, and $r_D$ represents
the size of the inner phase. $\chi_{\rm{in}}$ is the volume fraction of the inner phase,
i.e., $\chi_{\rm{in}}=\chi$ for droplet, rod, and slab configurations,
and $\chi_{\rm{in}}=1-\chi$ for tube and bubble configurations.
The charge density difference between the hadronic and quark phases is defined as
\begin{eqnarray}
\label{eq:dnc}
\delta n_c=n_c^{\rm{HP}}-n_c^{\rm{QP}},
\end{eqnarray}
with
\begin{eqnarray}
\label{eq:nch}
n_c^{\rm{HP}} &=& n_p - n_{e}^{\rm{HP}}-n_{\mu}^{\rm{HP}}, \\
n_c^{\rm{QP}} &=& \frac{2}{3}n_u -\frac{1}{3}n_d - \frac{1}{3}n_s - n_{e}^{\rm{QP}}-n_{\mu}^{\rm{QP}}.
\label{eq:ncq}
\end{eqnarray}
The surface tension $\sigma$ can significantly affect the structure of the mixed phase~\citep{Endo06,Yasu14,Wu19}.
At present, the value of $\sigma$ is poorly known, so it is usually taken as a free parameter.
In this study, we use the value of $\sigma=40$ MeV fm$^{-2}$; which is close to the
prediction of the MIT bag model using the multiple reflection expansion method~\citep{Ju21}.

The energy density of the mixed phase, given in Equation~(\ref{eq:fws}),
is calculated as a function of the following variables:
$n_{p}$, $n_{n}$, $n_{u}$, $n_{d}$, $n_{s}$, $n_{e}^{\rm{HP}}$, $n_{\mu}^{\rm{HP}}$,
$n_{e}^{\rm{QP}}$, $n_{\mu}^{\rm{QP}}$, $\chi$, and $r_D$.
The values of these variables are determined by solving a set of equilibrium equations
between the hadronic and quark phases at a given baryon density $n_b$.
In the EM method, the equilibrium conditions for coexisting two phases in the
cell are derived by minimizing the total energy density~(\ref{eq:fws})
under the constraints of baryon number conservation and globe charge neutrality,
which are written as
\begin{eqnarray}
\label{eq:nb}
  \frac{\chi}{3} \left(n_u+n_d+n_s\right)
  + \left(1 - \chi\right) \left( n_p + n_n \right) &=& n_b , \\
 \chi n_c^{\rm{QP}} + \left(1 - \chi\right) n_c^{\rm{HP}}  &=& 0  .
\label{eq:nc}
\end{eqnarray}
We introduce the Lagrange multipliers $\mu_n$ and $\mu_e$
for the constraints, and then perform the minimization for the function
\begin{eqnarray}
w &=& \varepsilon_{\rm{MP}}
  - \mu_n \left[ \frac{\chi}{3} \left(n_u+n_d+n_s\right)
  + \left(1 - \chi\right) \left( n_p + n_n \right) \right] \nonumber \\
&&+ \mu_e \left[ \chi n_c^{\rm{QP}} +\left(1 - \chi\right) n_c^{\rm{HP}} \right] .
\label{eq:w1}
\end{eqnarray}
According to the definition of the chemical potential,
$\mu_n=\partial \varepsilon_{\rm{MP}} / \partial n_n^{\rm{MP}} $ and
$\mu_e=\partial \varepsilon_{\rm{MP}} / \partial n_e^{\rm{MP}} $
correspond to the chemical potentials of neutrons and electrons in the mixed phase, respectively.
By minimizing $w$ with respect to the particle densities, we obtain
the following equilibrium conditions for chemical potentials:
\begin{eqnarray}
\mu_e &=& \mu_{e}^{\rm{HP}}-\frac{2\varepsilon_{\rm{Coul}}}{(1-\chi)\delta n_{c}}
        = \mu_{e}^{\rm{QP}}+\frac{2\varepsilon_{\rm{Coul}}}{\chi \delta n_{c}} ,
\label{eq:Ce1} \\
\mu_p &=& \mu_n - \mu_{e}^{\rm{HP}},
\label{eq:Cp1} \\
\mu_u &=& \frac{1}{3}\mu_n - \frac{2}{3}\mu_{e}^{\rm{QP}},
\label{eq:Cu1}\\
\mu_d &=& \mu_s = \frac{1}{3}\mu_n + \frac{1}{3}\mu_{e}^{\rm{QP}}.
\label{eq:Cd1}
\end{eqnarray}
It is necessary to clarify the definition of the chemical potential, especially for charged particles.
When the electric potential is taken into account, the chemical potential of a charged particle
is gauge dependent as discussed in~\citet{Tats03}. In the mixed phase studied here,
the hadronic and quark phases are charged separately, which lead to nonzero electric potentials
in the two phases. We can understand the terms associated with $\varepsilon_{\rm{Coul}}$ in
Equation~(\ref{eq:Ce1}) as the contribution from the electric potential.
The chemical potentials of nucleons and quarks in the above equations are defined by
Equations~(\ref{eq:mup}) and (\ref{eq:muq}), where the electric potential is not taken into account.
For the electrons, Equation~(\ref{eq:Ce1}) implies that the total chemical potential including
the contribution from the electric potential remains constant throughout the cell,
although the electron densities in the hadronic and quark phases are different from each other.
Moreover, Equations~(\ref{eq:Cp1})--(\ref{eq:Cd1}) imply that local
$\beta$ equilibrium should be reached, which is also satisfied in the Gibbs and Maxwell constructions.

The equilibrium condition for the pressures is derived by minimizing $w$
with respect to the volume fraction $\chi$, which is expressed as
\begin{eqnarray}
P_{\rm{HP}} &=& P_{\rm{QP}} -\frac{2\varepsilon_{\rm{Coul}}}{\delta n_c}
  \left[ \frac{n_c^{\rm{QP}}}{\chi} + \frac{n_c^{\rm{HP}}}{1-\chi}\right] \nonumber \\
& & \mp \frac{\varepsilon_{\rm{Coul}} }{\chi_{\rm{in}}}
  \left(3+\chi_{\rm{in}}\frac{{\Phi}^{^{\prime }}}{\Phi}\right),
\label{eq:CP1}
\end{eqnarray}
where the sign of the last term is \textquotedblleft $-$\textquotedblright\ for droplet,
rod, and slab configurations, while it is \textquotedblleft $+$\textquotedblright\ for
tube and bubble configurations. The pressure of the mixed phase is calculated from the
thermodynamic relation, $P_{\rm{MP}} = \mu_n  n_b - \varepsilon_{\rm{MP}}$,
which is actually equal to $-w$ given in Equation~(\ref{eq:w1}).
Comparing with the Gibbs condition of equal pressures,
the additional terms in Equation~(\ref{eq:CP1}) are caused by the finite-size effects.
When the surface and Coulomb energies are neglected by taking the limit
$\sigma \rightarrow 0$, all the equilibrium equations derived above
would reduce to the Gibbs conditions.

By minimizing $w$ with respect to the size $r_D$, we obtain the well-known
relation ${\varepsilon}_{\rm{surf}}=2{\varepsilon}_{\rm{Coul}}$, which leads to
the formula for the size of the inner phase,
\begin{eqnarray}
\label{eq:rd}
r_D &=& \left[\frac{\sigma{D}}{e^2\left(\delta n_c\right)^{2}\Phi}\right]^{1/3},
\end{eqnarray}
whereas the size of the Wigner--Seitz cell is given by
\begin{eqnarray}
\label{eq:rc}
r_C &=& \chi_{\rm{in}}^{-1/D} r_D.
\end{eqnarray}
In practice, we solve the equilibrium equations at a given baryon density $n_b$
for all pasta configurations, and then determine the thermodynamically stable state that
has the lowest energy density. The hadron-quark mixed phase exists only in the density
range where its energy density is lower than that of both hadronic matter and quark matter.

\begin{figure*}[htbp]
\begin{center}
\begin{tabular}{cc}
  \includegraphics[clip,width=8.0cm]{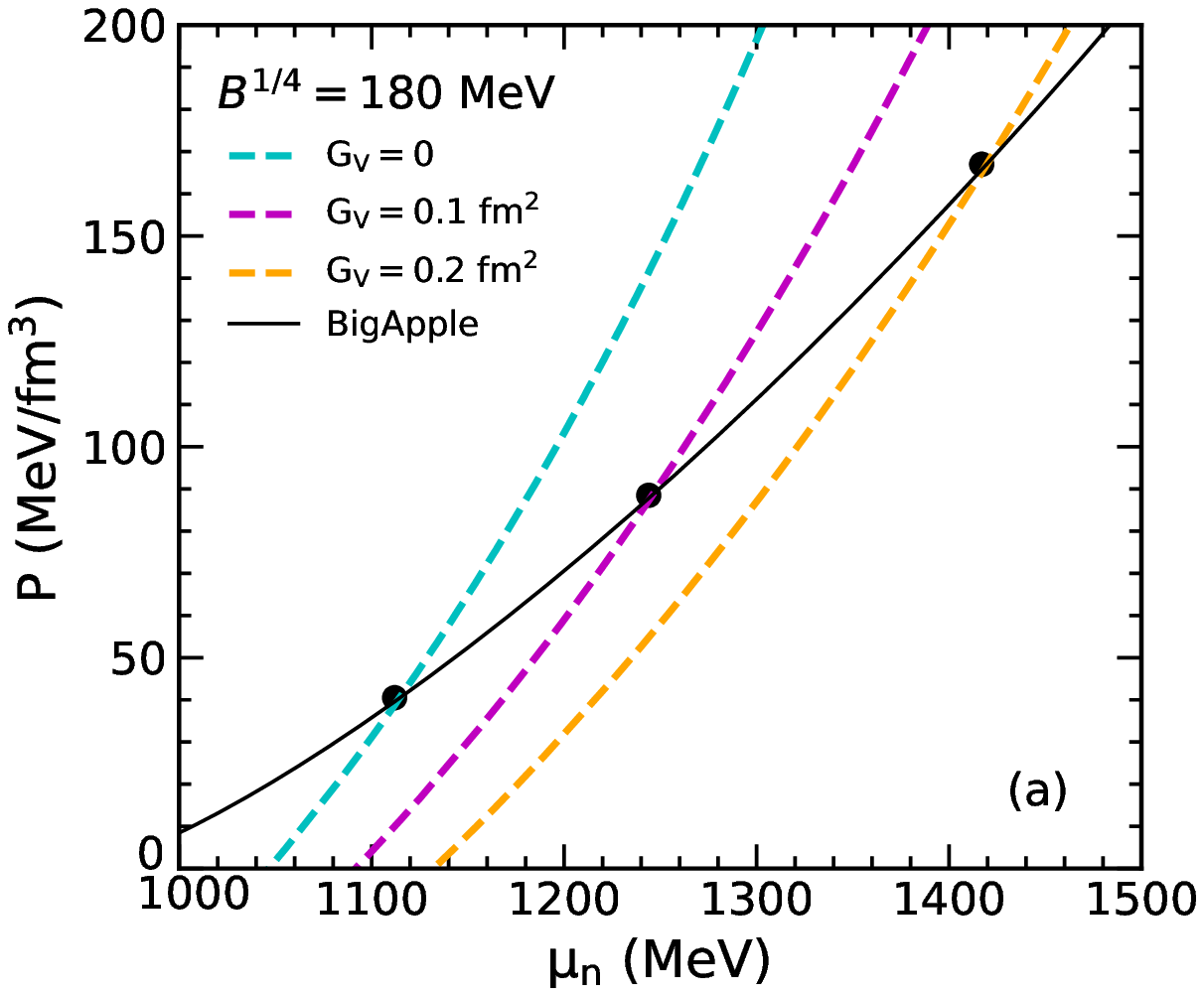} &
  \includegraphics[clip,width=8.0cm]{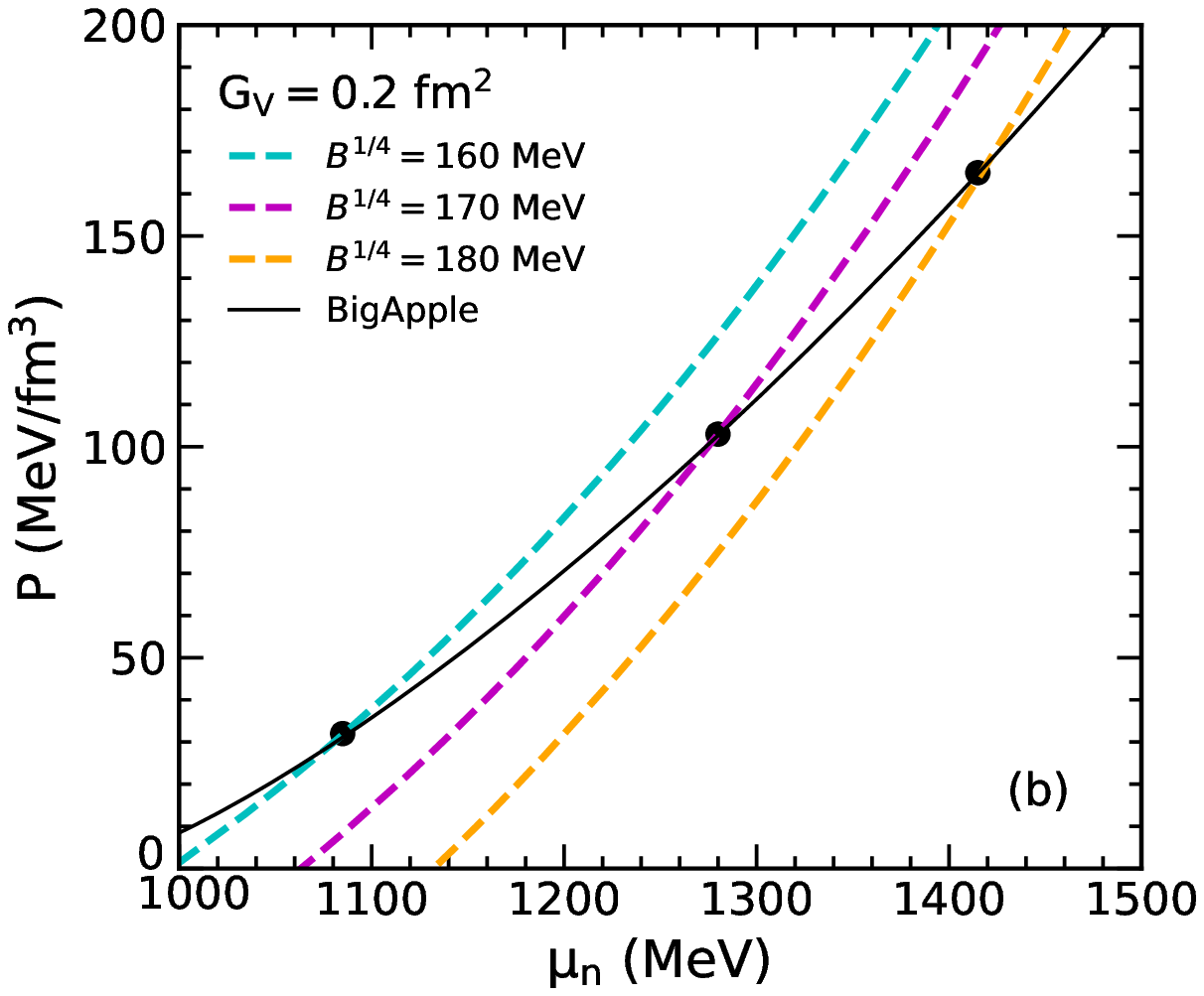} \\
\end{tabular}
  \caption{Pressure $P$ as a function of the neutron chemical potential $\mu_n$
  for the BigApple model and the vMIT model with different vector
  coupling $G_V$ (a) and different bag constant $B$ (b).
  The filled circles indicate the crossing points of the hadronic and quark
  curves corresponding to the phase transition in the Maxwell construction.}
  \label{fig:1pmun}
\end{center}
\end{figure*}

\begin{figure*}[htbp]
\begin{center}
\includegraphics[clip,width=17.0 cm]{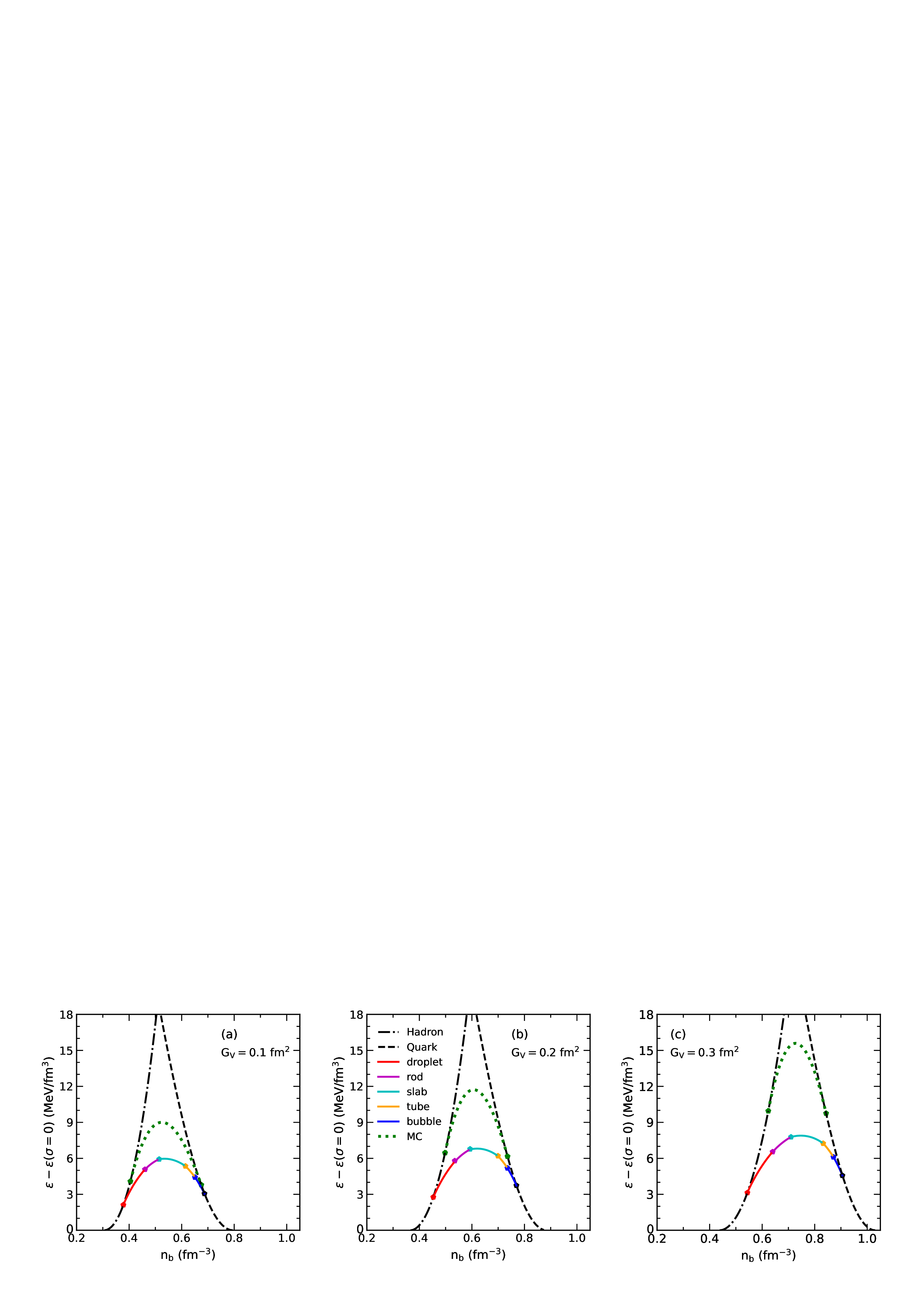}
\caption{Energy densities of the mixed phase obtained using the
EM method with $B^{1/4}=180$~MeV and $\sigma=40$~MeV fm$^{-2}$
relative to those of the GC ($\sigma=0$).
The filled circles indicate the transition points between different pasta phases.
The results of the Maxwell construction (MC) are shown by green dotted lines.}
\label{fig:2denb}
\end{center}
\end{figure*}

\begin{figure}[htbp]
\begin{center}
\includegraphics[clip,width=8.0 cm]{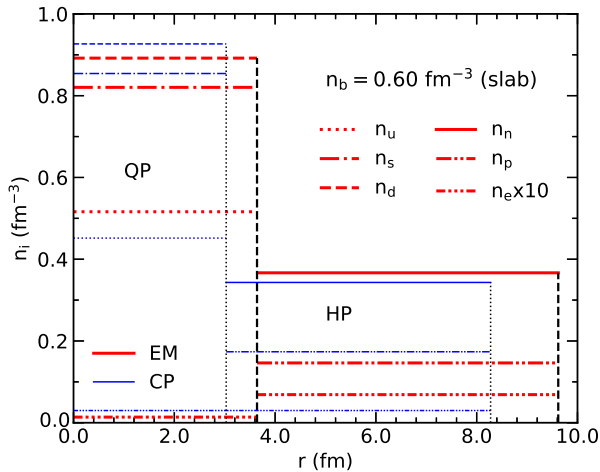}
\caption{Particle density profiles in the Wigner--Seitz cell for a slab
configuration at $n_{b}=0.6\, \rm{fm}^{-3}$.
The results are obtained with $B^{1/4}=180$~MeV and $G_V=0.2\, \rm{fm}^2$.
The hadron-quark interface and the cell boundary are indicated by the vertical lines.
The results of the EM method (red thick lines) are compared with those of
the CP method (blue thin lines) to show the charge screening effect.}
\label{fig:3nirws}
\end{center}
\end{figure}

\begin{figure*}[htbp]
\begin{center}
\includegraphics[clip,width=16.0 cm]{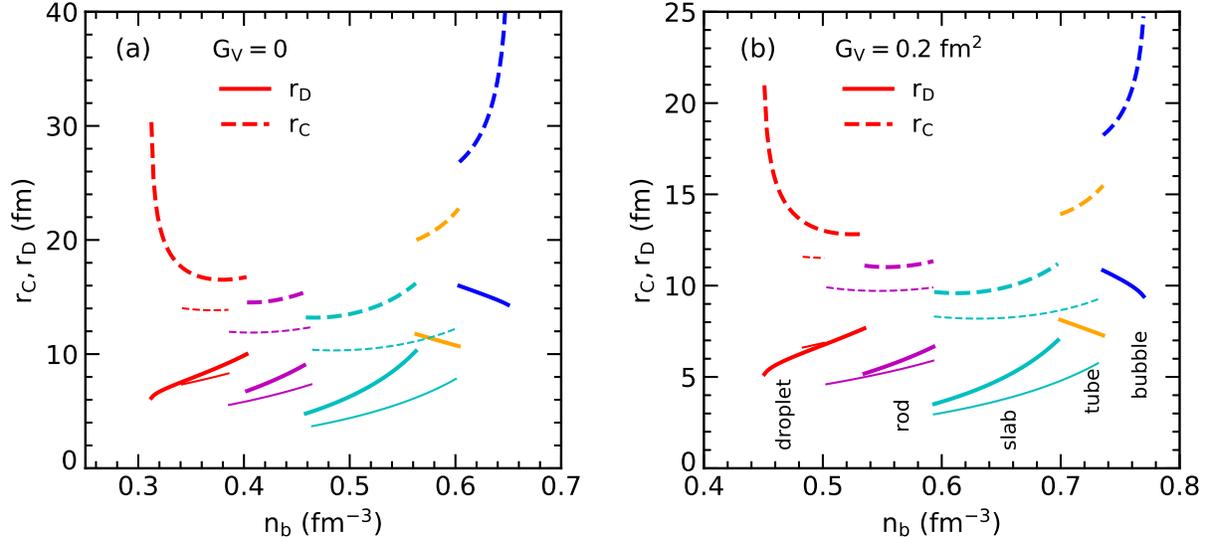}
\caption{Size of the Wigner--Seitz cell ($r_C$) and that of the inner
phase ($r_D$) as a function of the baryon density $n_b$ with $B^{1/4}=180$~MeV.
The results of the EM and CP methods are shown by thick and thin lines, respectively.}
\label{fig:4rnb}
\end{center}
\end{figure*}

\begin{figure}[htbp]
\begin{center}
\includegraphics[clip,width=8.0 cm]{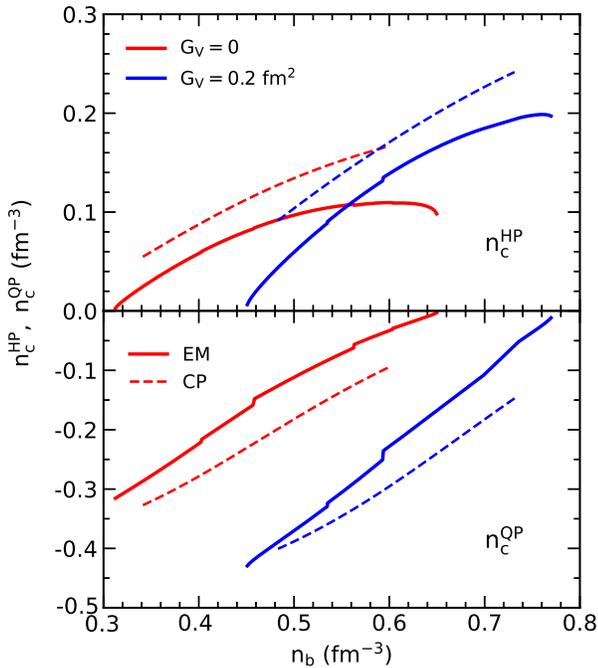}
\caption{Charge densities of hadronic and quark phases,
$n_c^{\rm{HP}}$ and $n_c^{\rm{QP}}$, as a function of
the baryon density $n_b$ with $B^{1/4}=180$~MeV.
The results of the EM and CP methods are shown by solid and dashed lines, respectively.}
\label{fig:5ncnb}
\end{center}
\end{figure}

\begin{table*}[htb]
\caption{Onset Densities of the Hadron-Quark Pasta Phases and Pure Quark
Matter Obtained Using Different Model Parameters and Methods.}
\label{tab:pasta}
\begin{center}
\begin{tabular}{ccccccccc}
\hline\hline
\multicolumn{2}{c}{Model Parameter}  &  Method   &  \multicolumn{6}{c}{Onset Density (fm$^{-3}$)}      \\
\cline{1-2} \cline{4-9}
$B^{1/4}$ (MeV)  &  $G_{V}$(fm$^{2}$)  &       &Droplet &Rod    &Slab   &Tube   &Bubble & Quark    \\
\hline
 180   & 0.1   & EM   &0.378  &0.460 &0.514 &0.615 &0.651 &0.688  \\
 180   & 0.2   & EM   &0.451  &0.535 &0.594 &0.700 &0.736 &0.770  \\
 180   & 0.3   & EM   &0.544  &0.640 &0.710 &0.832 &0.872 &0.907  \\
 170   & 0.2   & EM   &0.382  &0.442 &0.488 &0.571 &0.594 &0.609  \\
 170   & 0.2   & CP   &0.414  &0.419 &0.484 &  --  &  --  &0.580  \\
 180   & 0.2   & CP   &0.483  &0.503 &0.593 &  --  &  --  &0.732  \\
\hline\hline
\end{tabular}
\end{center}
\end{table*}

\begin{table*}[htbp]
\caption{Properties of Neutron Stars with the Maximum Mass $M_{\mathrm{max}}$.
}
\label{tab:nstar}
\begin{center}
\begin{tabular}{llccccccc}
\hline\hline
 \multicolumn{2}{c}{Model Parameter}  &  Method & $M_\mathrm{max}$ & $n_{b}^{c}$
 & $R_\mathrm{QP}$ & $R_\mathrm{MP}$ & $R$  \vspace{-0.05cm}\\
 \cline{1-2}
 $B^{1/4}$ (MeV) & $G_V$ $(\rm{fm}^{2})$  &   &$(M_\odot)$ & $(\rm{fm}^{-3})$
 & (km) & (km) & (km) \\
\hline
 $180$ & 0.1 & EM & 1.93 & 0.68 &  --  & 5.23 & 13.20 \\
 $180$ & 0.2 & EM & 2.32 & 0.69 &  --  & 5.27 & 13.10 \\
 $180$ & 0.3 & EM & 2.52 & 0.73 &  --  & 4.67 & 12.83 \\
 $170$ & 0.2 & EM & 2.08 & 0.76 & 4.54 & 7.00 & 12.65 \\
 $170$ & 0.2 & CP & 2.15 & 0.71 & 3.17 & 5.88 & 12.98 \\
 $180$ & 0.2 & CP & 2.36 & 0.70 &  --  & 4.46 & 13.05 \\
\hline\hline
\end{tabular}
\end{center}
\footnotesize{Note: The central baryon density is denoted by $n_{b}^{c}$, while
$R_{\mathrm{QP}}$, $R_{\mathrm{MP}}$, and $R$ correspond to the radii of
pure quark phase, mixed phase, and whole star, respectively.}
\end{table*}


\section{R\lowercase{esults and} D\lowercase{iscussion}}
\label{sec:4}
In this section, we present numerical results for the hadron-quark pasta phases,
which are likely to appear in the core of massive neutron stars.
We employ the RMF model with the BigApple parameterization to describe the hadronic matter.
For the quark matter, we employ the MIT bag model with vector interactions (vMIT),
while the effects of the vector coupling $G_V$ and the bag constant $B$ are discussed.
The hadron-quark mixed phases are computed by the EM method described in the
previous section, and the properties of massive neutron stars are calculated
by using the EOS with quarks.

\subsection{Hadron-quark mixed phases}
\label{sec:4-1}

In neutron-star matter, with increasing density the structured hadron-quark mixed phase
is expected to appear, which depends on the models used to describe the hadronic and
quark phases. In the present study, we employ the RMF model with the BigApple
parameterization for the hadronic phase, while the quark phase is described
by the vMIT model. To analyze the influence of model parameters on the deconfinement
phase transition, it is convenient to use the Maxwell construction in which the phase
transition appears at the crossing of the hadronic EOS with the quark EOS in the
pressure and chemical potential plane.
This is due to that two coexisting phases in the Maxwell construction must have
the same pressure and baryon chemical potential.
We show in Figure~\ref{fig:1pmun} the pressure $P$ as a function of the
neutron chemical potential $\mu_n$ for the BigApple model and the vMIT
model with different vector coupling $G_V$ (left panel)
and different bag constant $B$ (right panel).
According to the Maxwell equilibrium conditions, the phase transition takes place at
the crossing of the hadronic and quark EOS curves.
One can see from the left panel that a larger $G_V$ in the vMIT model corresponds to a
higher transition pressure, which implies that the deconfinement phase transition is
delayed accordingly. In the right panel, we see that the transition pressure increases
as the bag constant $B$ increases. Therefore, it is reasonable to expect that
the formation of hadron-quark pasta phases may be delayed with increasing vector
coupling $G_V$ and bag constant $B$ in the vMIT model.

We compute the properties of a structured hadron-quark mixed phase in the EM method,
where the surface and Coulomb energies are included in the minimization procedure.
Compared to the Gibbs calculation without finite-size effects, we obtain a higher
energy for the pasta phase, since the surface and Coulomb terms are always positive.
In Figure~\ref{fig:2denb}, we present the energy densities of pasta phases obtained
in the EM method with $B^{1/4}=180$~MeV and $\sigma=40$ MeV fm$^{-2}$
relative to those of the GC (i.e., $\sigma = 0$).
The filled circles represent the transition points between different pasta phases.
It is shown that the pasta configuration changes from droplet to rod, slab, tube,
and bubble with increasing baryon density $n_{b}$.
For comparison, the energy densities of pure hadronic matter and pure quark matter
are displayed by black dotted-dashed and dashed lines, respectively, whereas
the results of the Maxwell construction are shown by green dotted lines.
It is seen that the energy density of the Maxwell construction is higher than
that of the pasta phase, which implies that the structured mixed phase is
more stable than the Maxwell construction in our present calculations.
However, for large enough surface tension $\sigma$, the Maxwell construction
may have lower energy than the pasta phase, which has been discussed
in the literature~\citep{Masl19}.
To examine the influence of the quark vector interactions, we present the
results with $G_V=0.1$, 0.2, and 0.3 fm$^2$ in the left, middle, and right
panels, respectively. One can see that as $G_V$ increases,
the mixed phase is shifted toward higher densities with a wider range.
This tread is in agreement with the behavior observed in the Maxwell
construction (see the left panel of Figure~\ref{fig:1pmun}).
In Table~\ref{tab:pasta}, we present the onset densities of the hadron-quark pasta
phases and pure quark matter obtained using different model parameters and methods.
From this table, one can see the effect of $G_V$ and the difference between the
EM and CP methods.

It is interesting to study the geometric structure of the hadron-quark mixed phase,
which may be attributed to the competition between surface and Coulomb energies.
For checking the influence of the method used, we compare the results using the EM
method developed in the present work with those obtained in the simple CP method~\citep{Wu19}.
We emphasize that in the CP method the Gibbs conditions for phase
equilibrium are enforced,
while the surface and Coulomb energies are taken into account perturbatively,
and therefore the charge screening effect is disregarded in this case.
In contrast, the EM method incorporates the surface and Coulomb contributions
self-consistently in deriving the equilibrium conditions, which leads to
the rearrangement of charged particles known as the charge screening effect.
In order to see how large the charge screening effect is, we display
in Figure~\ref{fig:3nirws} the particle density profiles in the Wigner--Seitz cell
for a slab configuration at $n_{b}=0.6\, \rm{fm}^{-3}$.
The calculations are carried out with $B^{1/4}=180$~MeV and $G_V=0.2\, \rm{fm}^2$.
The hadron-quark interface and the cell boundary are indicated by the vertical lines.
One can see that the structure size of the EM method is larger than that of the CP method.
This is because the charge screening effect tends to reduce the net charge density
in each phase in order to lower the Coulomb energy.
In the negatively charged quark matter and positively charged hadronic matter,
the particle densities $n_{d}$, $n_{s}$, and $n_{p}$ are reduced in the EM method,
whereas $n_{u}$ is enhanced in comparison to that of the CP method.
The electron densities in the quark and hadronic matter are, respectively, reduced and
enhanced due to the same reason. It is expected that as the surface tension $\sigma$
increases, the charge screening effect becomes more pronounced, and finally the local
charge neutrality is reached in the Maxwell construction.

In Figure~\ref{fig:4rnb}, the size of the Wigner--Seitz cell ($r_C$) and that of
the inner phase ($r_D$) are plotted as a function of the baryon density $n_b$,
where the vMIT model with $B^{1/4}=180$~MeV are adopted for quark matter.
The results with $G_V=0$ and 0.2 fm$^2$ are shown in the left and right
panels, respectively. We compare the results obtained from the EM method (thick lines)
with those from the CP method (thin lines) in order to check the charge screening effect.
It is found that both $r_D$ and $r_C$ of the EM method are larger than that of the CP
method, which can be explained by the rearrangement of charged particles
as shown in Figure~\ref{fig:3nirws}.
According to Equation~(\ref{eq:rd}), a small charge density difference
$\delta n_c$ leads to a large pasta size $r_D$.
The value $\delta n_c=n_c^{\rm{HP}}-n_c^{\rm{QP}}$ can be observed in Figure~\ref{fig:5ncnb},
where the charge densities $n_c^{\rm{HP}}$ and $n_c^{\rm{QP}}$ are shown
as a function of the baryon density $n_b$.
One can see that the magnitudes of $n_c^{\rm{HP}}$ and $n_c^{\rm{QP}}$
in the EM method are significantly reduced due to the charge screening effect,
and hence their difference $\delta n_c$ is also reduced in comparison to the CP results.
The reduction of $\delta n_c$ caused by the charge rearrangement in the EM method
leads to the increase of $r_D$ and $r_C$ in Figures~\ref{fig:3nirws} and~\ref{fig:4rnb}.
In addition, a comparison between the EM and CP methods can be found in Table~\ref{tab:pasta},
where the onset and configuration of pasta phases are somewhat different.
As shown in the last two lines of Table~\ref{tab:pasta}, due to relatively large Coulomb energies
of the CP method, the tube and bubble configurations are energetically disfavored and
will not appear before the transition to pure quark matter.
A similar analysis is available for understanding the impact of the vector
coupling $G_V$ on the pasta size. Compared to the case of $G_V=0$ in the left panel
of Figure~\ref{fig:4rnb}, the values of $r_D$ and $r_C$ obtained with
$G_V=0.2\, \rm{fm}^2$ in the right panel are relatively small.
This can be understood from the behavior of $\delta n_c$ in Figure~\ref{fig:5ncnb},
where the values of $\delta n_c=n_c^{\rm{HP}}-n_c^{\rm{QP}}$ with $G_V=0.2\, \rm{fm}^2$
are clearly larger than those with $G_V=0$, which lead to smaller $r_D$ and $r_C$
in the right panel of Figure~\ref{fig:4rnb}.
Furthermore, we can see that as the density increases, $r_D$ in the droplet, rod, and slab
configurations increases, whereas it decreases in the tube and bubble phases.
These trends are related to a monotonic increase of the quark volume fraction $\chi$
during the phase transition.

\begin{figure*}[htbp]
\begin{center}
\includegraphics[clip,width=17.0 cm]{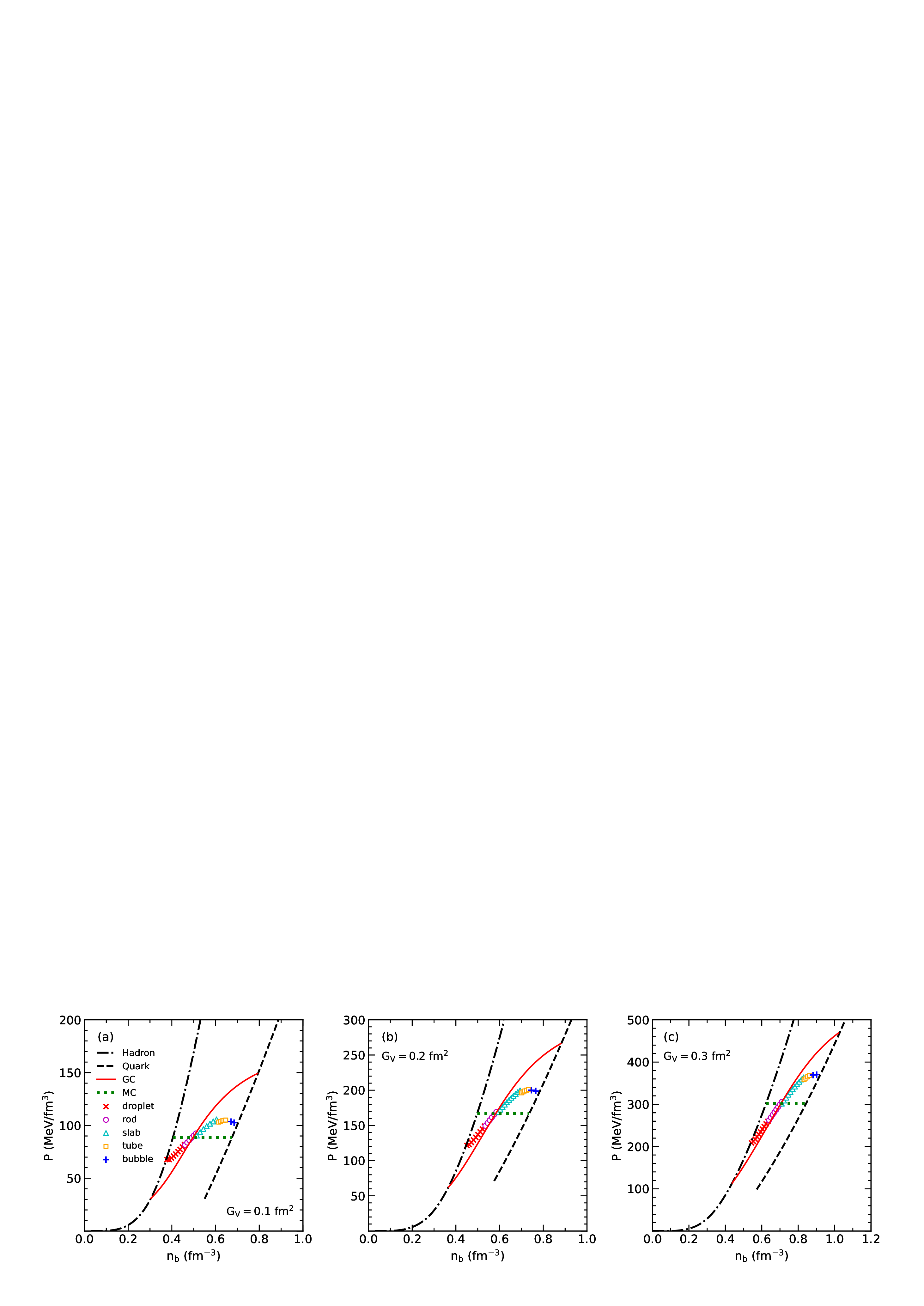}
\caption{Pressure $P$ as a function of the baryon density $n_b$ for hadronic, mixed,
and quark phases. The results of pasta phases obtained using
the EM method with $B^{1/4}=180$~MeV are compared to those of
the GC and MC.}
\label{fig:6pnb}
\end{center}
\end{figure*}

\begin{figure}[htbp]
\includegraphics[clip,width=8.0 cm]{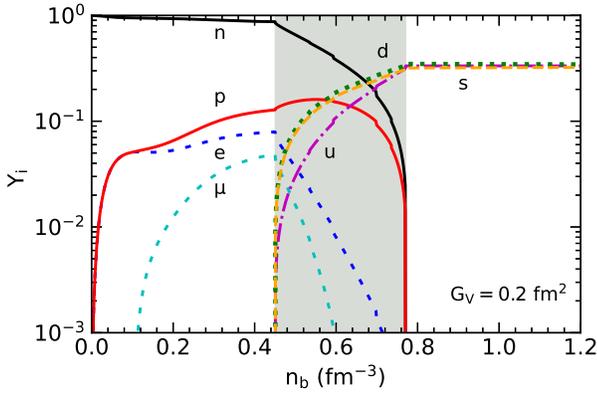}
\caption{Particle fraction $Y_i$ as a function of the baryon density $n_{b}$
for hadronic, mixed, and quark phases. The shaded area denotes the mixed phase
region, where the results of pasta phases are obtained using the EM method
with $B^{1/4}=180$~MeV and $G_V=0.2\, \rm{fm}^2$.}
\label{fig:7yinb}
\end{figure}

\begin{figure*}[htbp]
\begin{center}
\begin{tabular}{cc}
\includegraphics[clip,width=8.0cm]{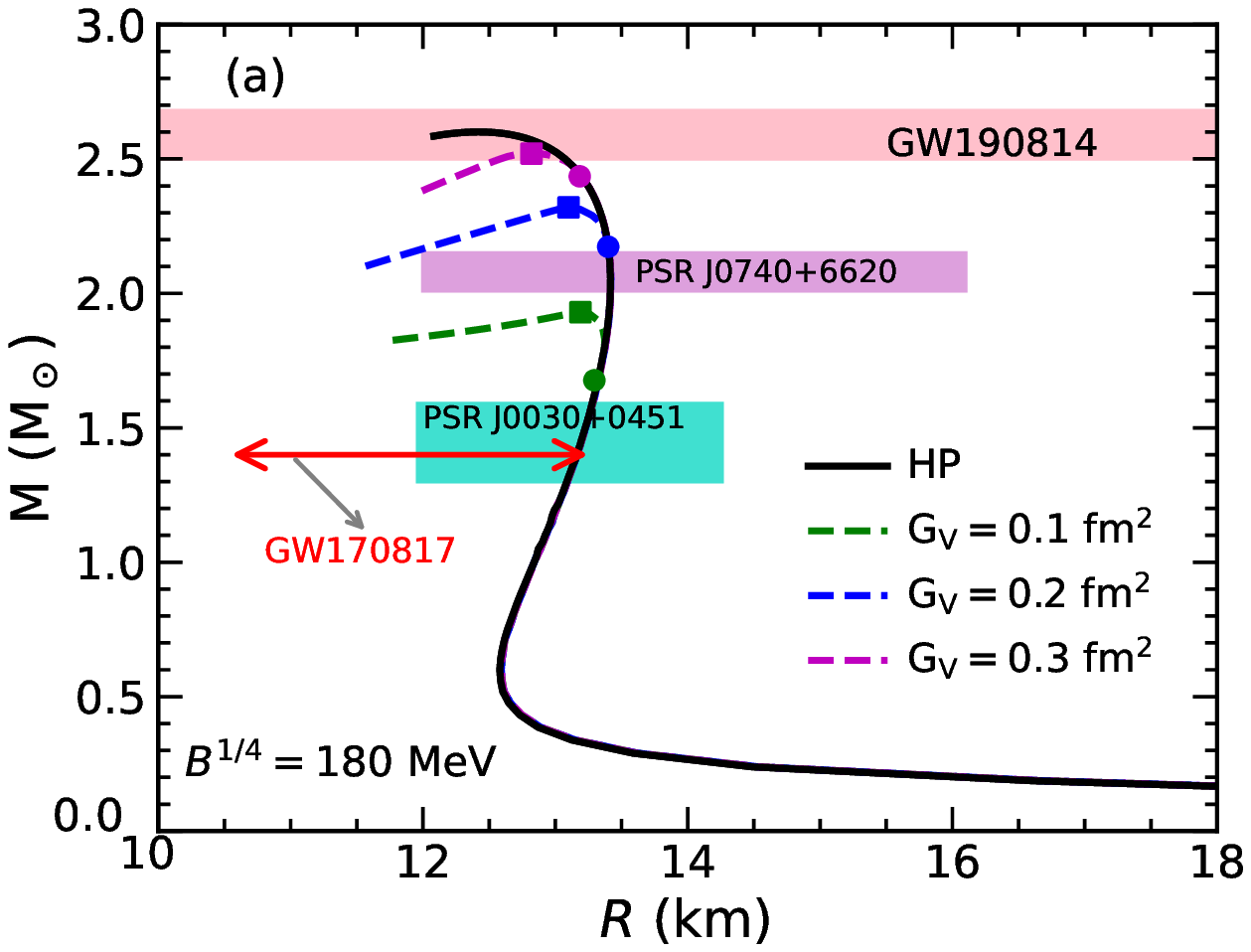} &
\includegraphics[clip,width=8.0cm]{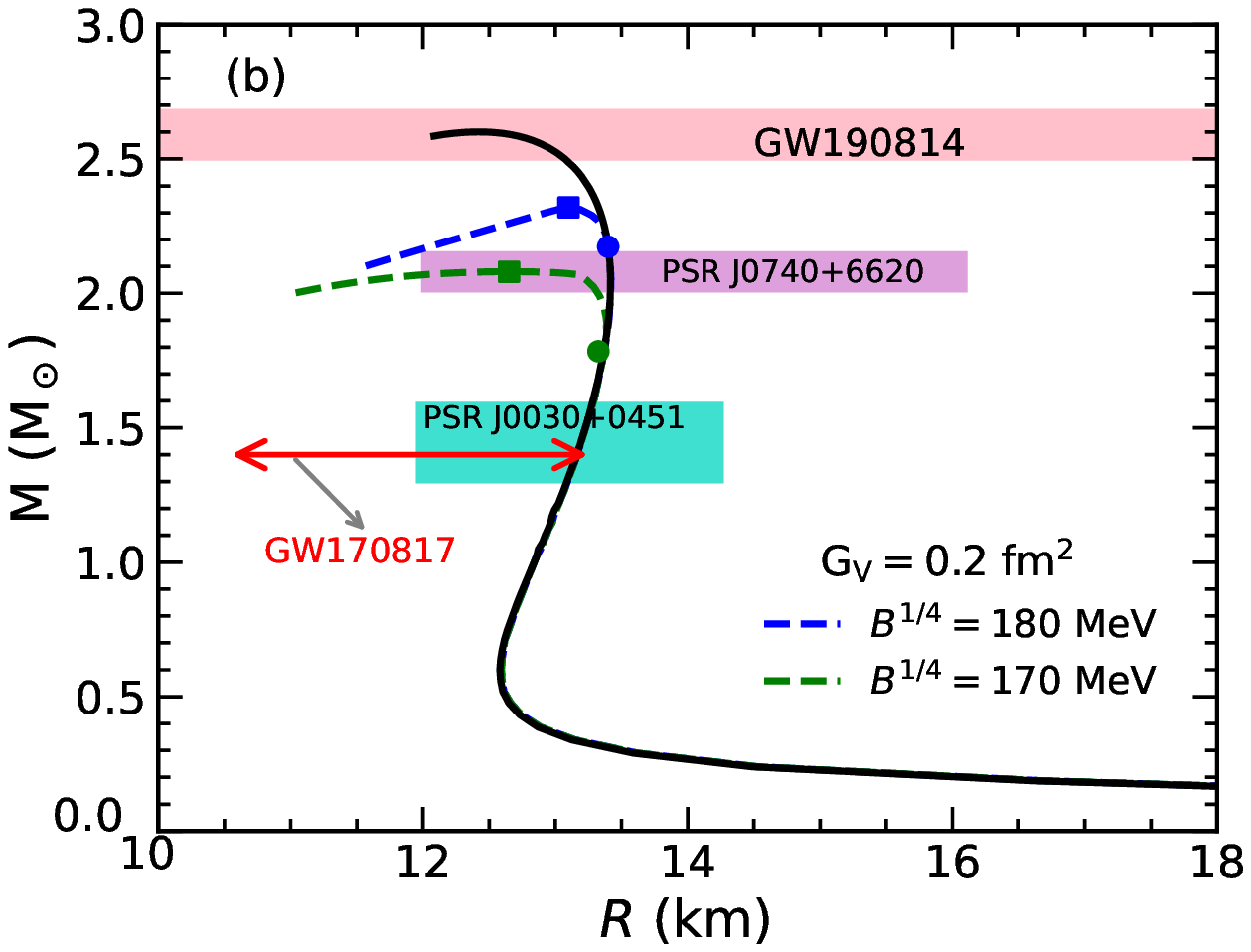} \\
\end{tabular}
\caption{Mass--radius relations of neutron stars for different model parameters.
The results of the pure hadronic EOS are compared to those including
the hadron-quark mixed phase described in the EM method for different vector
coupling $G_V$ (a) and different bag constant $B$ (b).
The filled circles indicate the onset of the star containing hadron-quark pasta phases,
while the filled squares denote the star with the maximum mass.
The red horizontal line with arrows at both ends represents the constraint on $R_{1.4}$
inferred from GW170817~\citep{Abbo18}.
The green and purple shaded areas correspond to simultaneous measurements of the mass and
radius from NICER for PSR J0030+0451~\citep{Mill19} and PSR J0740+6620~\citep{Mill21}, respectively.
The mass constraint from GW190814~\citep{Abbo190814} is depicted by the pink horizontal bar.}
\label{fig:8MR}
\end{center}
\end{figure*}

\begin{figure}[htbp]
\begin{center}
\includegraphics[clip,width=8.0 cm]{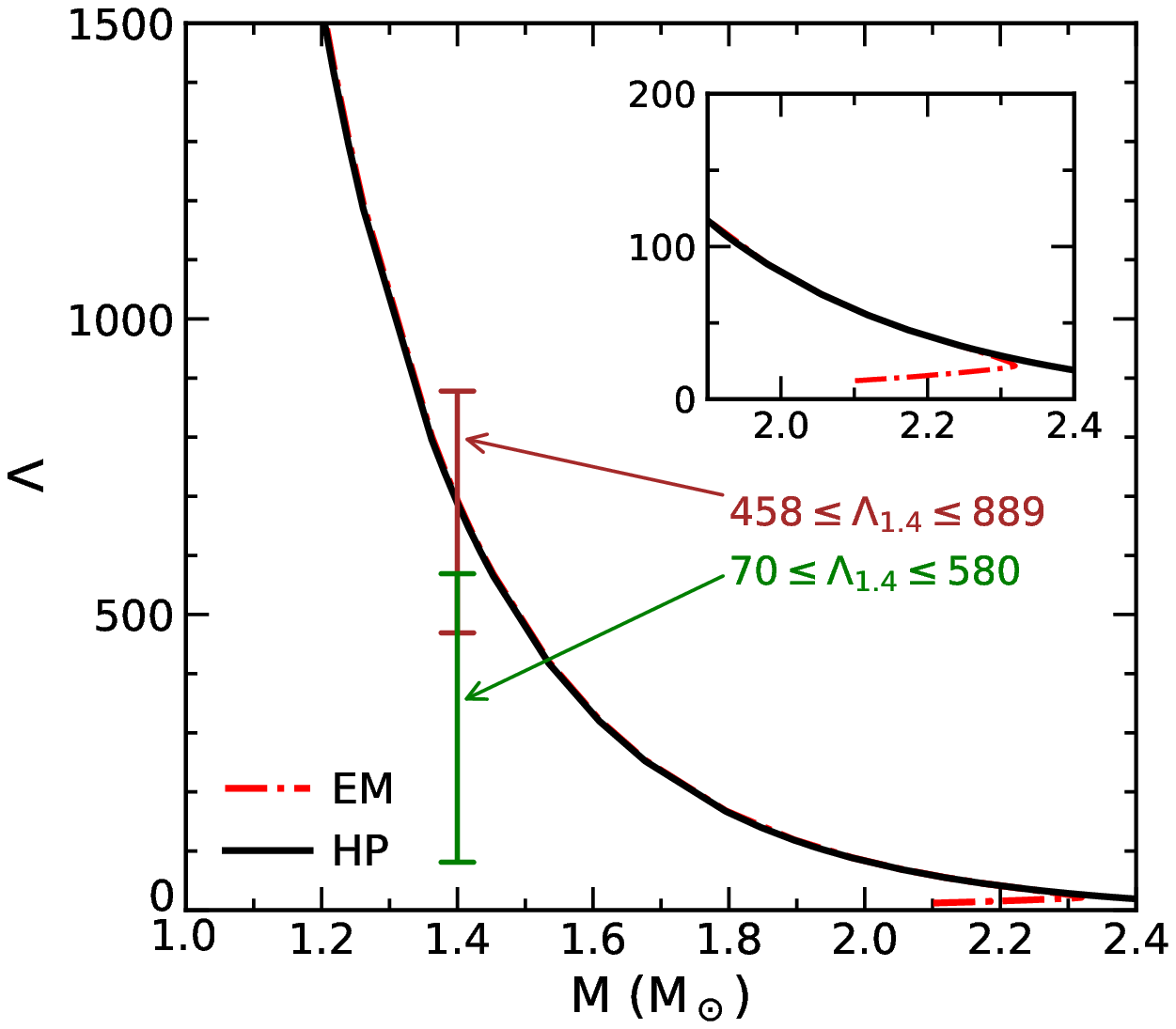}
\caption{Dimensionless tidal deformability $\Lambda$ as a function of the neutron-star mass $M$.
The results of the pure hadronic EOS are compared to those including the hadron-quark mixed phase
described in the EM method with $B^{1/4}=180$~MeV and $G_V=0.2$~fm$^2$.
The constraint $70 \le \Lambda_{1.4} \le 580$ is inferred from the analysis
of GW170817~\citep{Abbo18}, while $458 \le \Lambda_{1.4} \le 889$ is achieved
based on the assumption that GW190814's secondary is a neutron star~\citep{Abbo190814}. }
\label{fig:9ML}
\end{center}
\end{figure}


\subsection{Properties of \uppercase{n}eutron \uppercase{s}tars}
\label{sec:4-2}

In Figure~\ref{fig:6pnb}, we show the pressure $P$ as a function of the baryon
density $n_{b}$ for hadronic, mixed, and quark phases. The calculations of
hadron-quark pasta phases are performed in the EM method, where the hadronic
matter is described by the BigApple model and the quark matter by the vMIT model
with $B^{1/4}=180$~MeV.
For comparison, the results of the Gibbs
and Maxwell constructions are displayed by red solid and green dotted lines, respectively.
The pressures with the Maxwell construction remain constant during
the phase transition, whereas the pressures with the GC increase
with $n_{b}$ over a broad range.
It is shown that the results of pasta phases lie between the Gibbs and
Maxwell constructions. As the vector coupling $G_V$ increases from left to right panels,
we see that the hadron-quark mixed phases appear at higher densities and pressures.
Especially, the ending of the mixed phase shows a more significant $G_V$ dependence
than the beginning, which may be attributed to the increasing quark fraction
during the phase transition.

In Figure~\ref{fig:7yinb}, we display the particle fraction $Y_i=n_i/n_b$
as a function of the baryon density $n_{b}$.
The shaded area denotes the mixed phase region, where the results of hadron-quark
pasta phases are obtained in the EM method.
We employ the BigApple model for hadronic matter
and the vMIT model with $B^{1/4}=180$~MeV and $G_V=0.2\, \rm{fm}^2$ for quark matter.
At low densities, the matter consists of neutrons, protons, and electrons,
whereas the muons appear at about $0.11\, \rm{fm}^{-3}$.
When the deconfined quarks are present in the mixed phase,
the quark fractions $Y_u$, $Y_d$, and $Y_s$ increase rapidly together with a decrease of
neutron fraction $Y_n$. Meanwhile, $Y_e$ and $Y_\mu$ decrease significantly
because the quark matter is negatively charged that takes the role of electrons
for satisfying the constraint of global charge neutrality.
On the other hand, the hadronic matter in the mixed phase is positively charged,
so the proton fraction $Y_p$ increases slightly at the beginning of the mixed phase.
As the baryon density $n_{b}$ increases, $Y_p$ and $Y_n$ decrease to very low values
due to the increase of the quark volume fraction $\chi$ in the phase transition.
At sufficiently high densities, the matter turns into a pure quark phase,
where $Y_u \approx Y_d \approx Y_s \approx 1/3$ is nearly achieved under the
conditions of charge neutrality and chemical equilibrium.
The direct Urca (dUrca) process, known as the most efficient mechanism for neutron-star
cooling, is closely related to the particle fractions shown in Figure~\ref{fig:7yinb}.
In the hadronic matter, the dUrca process, i.e., the electron capture by a proton and
the $\beta$-decay of a neutron, is mainly determined by the proton fraction $Y_p$,
which must exceed a critical value to allow simultaneous energy and momentum conservation~\citep{Fant13}.
In the case of simple $npe$ matter, the dUrca process can occur only for $Y_p \geq 1/9$.
When muons are present under the equilibrium condition $\mu_e=\mu_\mu$,
the critical $Y_p$ for the dUrca process is in the range of $11.1\%-14.8\%$~\citep{Latt91b}.
According to observations of thermal radiation from neutron stars,
the dUrca process is unlikely to occur in neutron stars with masses below $1.5 M_\odot$,
since it would lead to an unacceptably fast cooling in disagreement with the observations~\citep{Fant13}.
In our present calculations, a $1.5 M_\odot$ neutron star
remains in the pure hadronic phase with a central density of about $0.34\, \rm{fm}^{-3}$,
where the proton fraction is lower than the critical value for the dUrca process.
We note that the BigApple model predicts a rather large threshold density of
$0.64\, \rm{fm}^{-3}$ for the dUrca process, which is related to its small
symmetry energy slope $L$ (see the discussion in~\citeauthor{Ji19} \citeyear{Ji19}).
This result is compatible with the cooling observations of neutron stars.

The properties of static neutron stars are obtained by solving the well-known
Tolman--Oppenheimer--Volkoff (TOV) equation with the EOS over a wide range of densities.
In the present calculations, we use the Baym--Pethick--Sutherland EOS~\citep{Baym71}
for the outer crust below the neutron drip density, while the inner crust EOS is based
on a Thomas--Fermi calculation using the TM1e parameterization of the RMF model~\citep{Shen20}.
The crust EOS is matched to the EOS of uniform neutron-star matter at the crossing
of the two segments. At very high densities, the quark degrees of freedom are
taken into account using the EM method for the hadron-quark pasta phases.
In Figure~\ref{fig:8MR}, we display the predicted mass-radius relations of
neutron stars, together with several constraints from astrophysical observations.
Compared to the results using pure hadronic EOS (solid lines),
the inclusion of quarks leads to an obvious reduction of the maximum
neutron-star mass $M_\mathrm{max}$. As shown in the left panel,
the results of massive stars are strongly dependent on the vector coupling $G_V$.
It is found that both the onset of hadron-quark pasta phases (filled circles) and
the values of $M_\mathrm{max}$ (filled squares) increase with increasing $G_V$.

 For a canonical $1.4 M_\odot$ neutron star, only the pure hadronic phase is present.
The prediction of the BigApple model for the radius $R_{1.4}$ is compatible
with the constraints inferred from the GW170817 event and NICER for PSR J0030+0451.

In order to examine the impact of the bag constant $B$, we compare in the right panel
the results between $B^{1/4}=180$~MeV (blue dashed line) and $B^{1/4}=170$~MeV (green dashed line).
The reduction of $M_\mathrm{max}$ with $B^{1/4}=170$~MeV is more pronounced than
with $B^{1/4}=180$~MeV. This is because the formation of hadron-quark pasta phases
appears at lower densities for $B^{1/4}=170$~MeV (see Table~\ref{tab:pasta}),
which leads to a stronger softening of the EOS.
To analyze the effects of the hadron-quark phase transition in more detail, we show
in Table~\ref{tab:nstar} the resulting properties of neutron stars with the maximum mass.
It is found that both $G_V$ and $B$ can significantly affect the structure and maximum mass
of neutron stars, whereas the influence of the charge screening (i.e., the difference between the
EM and CP methods) is relatively small.
In most cases, a structured hadron-quark mixed phase can be formed in massive neutron stars
inside the radius $R_{\mathrm{MP}}$. The pure quark matter appears in the interior of
neutron stars only for the cases with small $B$ where the central
density $n_{b}^{c}$ is larger than the onset of the pure quark phase.
We find that for all cases listed in Table~\ref{tab:nstar}, the central density of
a canonical $1.4 M_\odot$ neutron star is not high enough to form hadron-quark pasta
phases, so it remains in a pure hadronic phase.
These results are qualitatively consistent with the arguments of~\citet{Anna20},
which suggested that the matter in the core of a $1.4 M_\odot$ neutron star
is compatible with nuclear model calculations, whereas the matter in the interior
of a $2 M_\odot$ neutron star exhibits characteristics of the deconfined quark phase.
In Figure~\ref{fig:9ML}, we plot the dimensionless tidal deformability $\Lambda$ as a function
of the neutron-star mass $M$. The results of the pure hadronic EOS
are compared to those including the hadron-quark mixed phase described in the EM method
with $B^{1/4}=180$~MeV and $G_V=0.2$~fm$^2$.
It is shown that the inclusion of quarks leads to small differences in $\Lambda$ for
massive neutron stars, which is related to the reduction of the radius as shown in Figure~\ref{fig:8MR}.
Considering the constraints on $\Lambda_{1.4}$ inferred from the analysis of GW170817
and GW190814, the BigApple model provides acceptable results for the tidal deformability
of neutron stars. The effects of the hadron-quark phase transition can be observed only
for very massive stars as shown in the inset of Figure~\ref{fig:9ML}.

\section{S\lowercase{ummary}}
\label{sec:5}

Motivated by the recent advances in astrophysical observations, we studied the
properties of the hadron-quark pasta phases, which may appear in the interior
of massive neutron stars. The structured mixed phase is described
within the Wigner--Seitz approximation, where the whole space is divided into
equivalent cells with a geometric symmetry. The coexisting hadronic and quark
phases inside the cell are assumed to be separated by a sharp interface
with constant densities in each phase.
We extended the EM method for describing the hadron-quark pasta phases
by allowing different electron densities in the two coexisting phases,
which is helpful for understanding the transition from the GC
to the MC. In the EM method, the surface and Coulomb energies
are included in the minimization procedure, which results in different
equilibrium equations from the Gibbs conditions.
Comparing to the simple CP method, the treatment of surface and Coulomb
energies in the EM method leads to the rearrangement of charged particles,
known as the charge screening effect, which can significantly affect
the structure of the hadron-quark mixed phase.
The resulting EOS obtained from the EM method was found to lie between those
of the GC and MC.

In the present study, we have employed the RMF model with the BigApple
parameterization to describe the hadronic matter, while the quark matter
is described by a modified MIT bag model with vector interactions (vMIT).
The BigApple model could provide a good description of finite nuclei across
the nuclear chart, while its prediction for the maximum neutron-star mass
is as large as $2.6 M_\odot$. In addition, the BigApple model predicts
acceptable radius and tidal deformability for a canonical $1.4 M_\odot$ neutron
star, as compared to the estimations from astrophysical observations.
For the quark matter in the vMIT model, the vector interactions among quarks
could significantly stiffen the EOS at high densities and help to enhance
the maximum mass of neutron stars. We found that as the vector coupling $G_V$
increases, the hadron-quark pasta phases appear at higher densities,
and especially the ending of the mixed phase shows a more significant $G_V$
dependence than the beginning.
In addition, a larger $G_V$ leads to a smaller pasta size, which is caused
by relatively large charge density. Meanwhile, the pasta size is also affected
by the charge screening effect in the EM method, where the rearrangement
of charged particles can lower the Coulomb energy and enhance the pasta size.

We investigated the properties of massive neutron stars containing the hadron-quark
pasta phases. It was found that the inclusion of quarks could
considerably soften the EOS and reduce the maximum mass of neutron stars.
The results of massive stars clearly depend on the vector coupling $G_V$
and the bag constant $B$ used. A large $G_V$ corresponds to a weak reduction of
$M_\mathrm{max}$ relative to that of pure hadronic stars.
In most cases considered in this study, a structured hadron-quark
mixed phase can be formed in the interior of massive neutron stars,
whereas a canonical $1.4 M_\odot$ neutron star remains in a pure hadronic phase.
Our results showed that the presence of quarks inside neutron stars could be
compatible with current constraints inferred from NICER data and
gravitational-wave observations.

\acknowledgments

This work was supported in part by the National Natural Science Foundation of
China (grants Nos. 12175109 and 11775119).



\begin{thebibliography}{}
\bibitem[Abbott et al.(2017)]{Abbo17}
Abbott, B. P., Abbott, R., Abbott, T. D., et al.
2017, \prl, 119, 161101
	
\bibitem[Abbott et al.(2018)]{Abbo18}
Abbott, B. P., Abbott, R., Abbott, T. D., et al.
2018, \prl, 121, 161101

\bibitem[Abbott et al.(2020a)]{Abbo190425}
Abbott, B. P., Abbott, R., Abbott, T. D., et al.
2020, \apjl, 892, L3

\bibitem[Abbott et al.(2020b)]{Abbo190814}
Abbott, B. P., Abbott, R., Abbott, T. D., et al.
2020, \apjl, 896, L44

\bibitem[Alam et al.(2016)]{Alam16}
Alam, N., Agrawal, B.~K., Fortin, M., Pais, H., Provid{\^{e}}ncia, C., Raduta, Ad.~R., \& Sulaksono, A.
2016, \prc, 94, 052801(R)

\bibitem[Annala et al.(2020)]{Anna20}
Annala, E., Gorda, T., Kurkela, A., N\"attil\"a, J., \& Vuorinen, A.
2020, NatPh, 16, 907

\bibitem[Antoniadis et al.(2013)]{Anto13}
Antoniadis, J., Freire, P. C. C., Wex, N., et al.
2013, Sci, 340, 448

\bibitem[Arzoumanian et al.(2018)]{Arzo18}
Arzoumanian, Z., et al. 2018, \apjs, 235, 37

\bibitem[Bao et al.(2014)]{Bao14}
Bao, S. S., Hu, J. N., Zhang, Z. W., \& Shen, H.
2014, \prc, 90, 045802

\bibitem[Baym et al.(1971)]{Baym71}
Baym, G., Bethe, H. A., \& Pethick, C. J.
1971, \nphysa, 175, 225

\bibitem[Baym et al.(2018)]{Baym18}
Baym, G., Hatsuda, T., Kojo, T., Powell, P. D., Song, Y., \& Takatsuka, T.
2018, RPPh, 81, 056902

\bibitem[Bhattacharyya et al.(2010)]{Bhat10}
Bhattacharyya, A., Mishustin, I. N., \& Greiner, W.
2010, JPhG, 37, 025201

\bibitem[Cromartie et al.(2020)]{Crom20}
Cromartie, H. T., Fonseca, E., Ransom, S. M., et al.
2020, NaAs, 4, 72

\bibitem[Demircik et al.(2021)]{Demi21}
Demircik, T., Ecker, C., \& J\"arvinen, M.
2021, \apjl, 907, L37

\bibitem[Demorest et al.(2010)]{Demo10}
Demorest, P. B., Pennucci, T., Ransom, S. M., Roberts, M. S. E., \& Hessels, J. W. T.
2010, Natur, 467, 1081

\bibitem[Dexheimer et al.(2021)]{Dexh21}
Dexheimer, V., Gomes, R., Kl\"ahn, T., Han, S., \& Salinas, M.
2021, \prc, 103, 025808

\bibitem[Endo et al.(2006)]{Endo06}
Endo, T., Maruyama, T., Chiba, S., \& Tatsumi, T.
2006, PThPh, 115, 337

\bibitem[Essick \& Landry(2020)]{Essi20}
Essick, R., \& Landry, P.
2020, \apj, 904, 80

\bibitem[Fantina et al.(2013)]{Fant13}
Fantina, A. F., Chamel, N., Pearson, J. M., \& Goriely, S.
2013, A\&A, 559, A128

\bibitem[Fattoyev et al.(2020)]{Fatt20}
Fattoyev, F. J., Horowitz, C. J., Piekarewicz, J., \& Reed, B.
2020, \prc, 102, 065805

\bibitem[Fattoyev et al.(2010)]{Fatt10}
Fattoyev, F. J., Horowitz, C. J., Piekarewicz, J., \& Shen, G.
2010, \prc, 82, 055802

\bibitem[Fonseca et al.(2016)]{Fons16}
Fonseca, E., et al. 2016, \apj, 832, 167

\bibitem[Fonseca et al.(2020)]{Fons21}
Fonseca, E., et al. 2021, \apjl, 915, L12

\bibitem[Glendenning(1992)]{Glen92}
Glendenning, N. K.
1992, \prd, 46, 1274

\bibitem[Gomes et al.(2019)]{Gome19}
Gomes, R. O., Char, P., \& Schramm, S.
2019, \apj, 877, 139

\bibitem[Han et al.(2019)]{Han19}
Han, S., Mamun, M. A. A., Lalit, S., Constantinou, C., \& Prakash, M.
2019, \prd, 100, 103022

\bibitem[Heiselberg et al.(1993)]{Heis93}
Heiselberg, H., Pethick, C. J., \& Staubo, E. F.
1993, \prl, 70, 1355

\bibitem[Huang et al.(2020)]{Huang20}
Huang, K. X., Hu, J. N., Zhang, Y., \& Shen, H.
2020, \apj, 904, 39

\bibitem[Ji et al.(2019)]{Ji19}
Ji, F., Hu, J. N., Bao, S. S., \& Shen, H.
2019, \prc, 100, 045801

\bibitem[Ju et al.(2021)]{Ju21}
Ju, M., Wu, X. H., Ji, F., Hu, J. N., \& Shen, H.
2021, \prc, 103, 025809

\bibitem[Kl\"{a}hn \& Fischer (2015)]{Klah15}
Kl\"{a}hn, T. \& Fischer, T.
2015, \apj, 810, 134

\bibitem[Lalazissis et al.(1997)]{Lala97}
Lalazissis, G. A., K\"oning, J., \& Ring, P.
1997, \prc, 55, 540

\bibitem[Lattimer \& Swesty(1991)]{Latt91}
Lattimer,~J.~M., \& Swesty,~F.~D.
1991, \nphysa, 535, 331

\bibitem[Lattimer et al.(1991)]{Latt91b}
Lattimer, J. M., Pethick, C. J., Prakash, M., \& Haensel, P.
1991, \prl, 66, 2701

\bibitem[Lattimer \& Prakash(2016)]{Latt16}
Lattimer, J. M., \& Prakash, M.
2016, PhR, 621, 127

\bibitem[Li et al.(2020)]{Li20}
Li, J. J., Sedrakian, A., \& Weber, F.
2020, PhLB, 810, 135812

\bibitem[Lonardoni et al.(2015)]{Lona15}
Lonardoni, D., Lovato, A., Gandolfi, S., \& Pederiva, F.
2015, \prl, 114, 092301

\bibitem[Lopes et al.(2021)]{Lope21}
Lopes, L., L., Biesdorf, C., \& Menezes, D. P.
2021, PhyS, 96, 065303

\bibitem[Maruyama et al.(2007)]{Maru07}
Maruyama, T., Chiba, S., Schulze, H.-J., \& Tatsumi, T.
2007, \prd, 76, 123015

\bibitem[Maslov et al.(2019)]{Masl19}
Maslov, K., Yasutake, N., Blaschke, D., Ayriyan, A., Grigorian, H.,
Maruyama, T., Tatsumi, T., \& Voskresensky, D. N.
2019, \prc, 100, 025802

\bibitem[Miller et al.(2019)]{Mill19}
Miller, M. C., Lamb, F. K., Dittmann, A. J., et al.
2019, \apjl, 887, L24

\bibitem[Miller et al.(2021)]{Mill21}
Miller, M. C., Lamb, F. K., Dittmann, A. J., et al.
2021, \apjl, 918, L28	

\bibitem[Most et al.(2020)]{Most20}
Most, E. R., Papenfort, L. J., Weih, L. R., \& Rezzolla, L.
2020, MNRAS Lett, 499, L82

\bibitem[Oertel et al.(2017)]{Oert17}
Oertel, M., Hempel, M., Kl\"{a}hn, T., \& Typel, S.
2017, RvMP, 89, 015007

\bibitem[Riley et al.(2019)]{Rile19}
Riley, T. E., Watts, A. L., Bogdanov, S., et al.
2019, \apjl, 887, L21

\bibitem[Riley et al.(2021)]{Rile21}
Riley, T. E., Watts, A. L., Ray, P. S., et al.
2021, ApJL, 918, L27

\bibitem[Schertler et al.(2000)]{Sche00}
Schertler, K., Greiner, C., Schaffner-Bielich, J., \& Thoma, M. H.
2000, \nphysa, 677, 463

\bibitem[Shen et al.(2020)]{Shen20}
Shen, H., Fan, J., Hu, J. N., \& Sumiyoshi, K.
2020, \apj, 891, 148

\bibitem[Spinella et al.(2016)]{Spin16}
Spinella, W. M., Weber, F., Contrera, G. A., \& Orsaria, M. G.
2016, EPJA, 52, 61

\bibitem[Sugahara \& Toki(1994)]{Suga94}
Sugahara, Y., \& Toki, H.
1994, \nphysa, 579, 557

\bibitem[Tan et al.(2020)]{Tan20}
Tan, H., Noronha-Hostler, J., \& Yunes, N.
2020, \prl, 125, 261104

\bibitem[Tatsumi et al.(2003)]{Tats03}
Tatsumi, T., Yasuhira, M., \& Voskresensky, D.
2003, \nphysa, 718, 359

\bibitem[Tews et al.(2021)]{Tews21}
Tews, I., Pang, P. T. H., Dietrich, T., Coughlin, M. W., Antier, S.,
Bulla, M., Heinzel, J., \& Issa, L.
2021, \apjl, 908, L1

\bibitem[Tsokaros et al.(2020)]{Tsok20}
Tsokaros, A., Ruiz, M., \& Shapiro, S. L.
2020, \apj, 905, 48

\bibitem[Weber et al.(2019)]{Webe19}
Weber, F., Farrell, D., Spinella, W. M., Malfatti, G., Orsaria, M. G.,
Contrera, G. A., \& Maloney, I.
2019, Univ, 5, 169

\bibitem[Wu \& Shen(2017)]{Wu17}
Wu, X. H. \& Shen, H.
2017, \prc, 96, 025802

\bibitem[Wu \& Shen(2019)]{Wu19}
Wu, X. H. \& Shen, H.
2019, \prc, 99, 065802

\bibitem[Xu et al.(2010)]{Xu10}
Xu, J., Chen, L. W., Ko, C. M., \& Li, B. A.
2010, \prc, 81, 055803

\bibitem[Yang \& Shen(2008)]{Yang08}
Yang, F. \& Shen, H.
2008, \prc, 77, 025801

\bibitem[Yasutake et al.(2014)]{Yasu14}
Yasutake, N., {\L}astowiecki, R., Beni{\'{c}}, S., Blaschke, D.,
Maruyama, T., \& Tatsumi, T.
2014, \prc, 89, 065803

\bibitem[Zhang \& Li(2020)]{Zhang20}
Zhang, N. B., \& Li, B. A.
2020, \apj, 902, 38

\end{thebibliography}
\end{document}